\documentclass[epjCONF]{svjour}
\usepackage{epsfig}
\usepackage{subfigure}
\usepackage{graphics}
\usepackage[varg]{txfonts} % Times fonts
\usepackage[latin1]{inputenc}
\session-title{Hot and Cold Baryonic Matter -- HCBM 2010}
\begin{document}
\title{Dynamical equilibration in strongly-interacting parton-hadron matter}
\author{Vitalii Ozvenchuk \inst{1}\fnmsep\thanks{\email{ozvenchuk@fias.uni-frankfurt.de}}
\and Elena Bratkovskaya \inst{1,2} \and Olena Linnyk \inst{2} \and
Mark Gorenstein \inst{3} \and Wolfgang Cassing \inst{4} }
\institute{Frankfurt Institute for Advanced Studies, Frankfurt am
Main, Germany \and Institut f\"ur Theoretische Physik, Goethe
Universit\"at, Frankfurt am Main, Germany  \and Bogolyubov Institute
for Theoretical Physics, Kiev, Ukraine \and Institut f\"ur
Theoretische Physik, Justus Liebig Universit\"at, Gie\ss en,
Germany}
\abstract{ We study the kinetic and chemical equilibration in
`infinite' parton-hadron matter within the Parton-Hadron-String
Dynamics transport approach, which is based on a dynamical
quasiparticle model for partons matched to reproduce lattice-QCD
results -- including the partonic equation of state -- in
thermodynamic equilibrium. The `infinite' matter is simulated within
a cubic box with periodic boundary conditions initialized at
different baryon density (or chemical potential) and energy density.
The transition from initially pure partonic matter to hadronic
degrees of freedom (or vice versa) occurs dynamically by
interactions. Different thermodynamical distributions of the
strongly-interacting quark-gluon plasma (sQGP) are addressed and
discussed.
} %end of abstract
\maketitle

\section{Introduction}
\label{intro}

Nucleus-nucleus collisions at ultra-relativistic energies are
studied experimentally and theoretically to obtain information about
the properties of hadrons at high density and/or temperature as well
as  about the phase transition to a new state of matter, the
quark-gluon plasma (QGP). Whereas the early `big-bang' of the
universe most likely evolved through steps of kinetic and chemical
equilibrium, the laboratory `tiny bangs' proceed through phase-space
configurations that initially are far from an equilibrium phase and
then evolve by fast expansion. On the other hand, many observables
from strongly-interacting systems are dominated by many-body phase
space such that spectra and abundances look `thermal'.  It is thus
tempting to characterize the experimental observables by global
thermodynamical quantities like 'temperature', chemical potentials
or entropy \cite{Ref1,Ref2,Ref3,Ref4,Ref5,Ref6,Ref7,Ref8}. We note,
that the use of macroscopic models like hydrodynamics
\cite{Ref9,Ref10,Ref11,Ref12} employs as basic assumption the
concept of local thermal and chemical equilibrium. The crucial
question, however, how and on what timescales a global thermodynamic
equilibrium can be achieved, is presently a matter of debate. Thus
nonequilibrium approaches have been used in the past to address the
problem of timescales associated to global or local equilibration
\cite{Ref13,Ref14,Ref15,Ref16,Ref17,Ref18,Ref19,Ref20,Ref21}.  In
view of the increasing `popularity' of thermodynamic analyses a
thorough microscopic study of the questions of thermalization and
equilibration of confined and deconfined matter within a transport
approach appears necessary.

\section{The model}
\label{sec:model}

In this contribution, we study the kinetic and chemical
equilibration in `infinite' parton-hadron matter within the novel
Parton-Hadron-String Dynamics (PHSD) transport approach
\cite{Ref22,Ref23}, which is based on generalized transport
equations on the basis of the off-shell Kadanoff-Baym equations
\cite{Ref24,Ref25} for Green's functions in phase-space
representation (in the first order gradient expansion, beyond the
quasiparticle approximation). In the KB theory, the field quanta are
described in terms of propagators with complex self-energies.
Whereas the real part of the self-energies can be related to
mean-field potentials, the imaginary parts provide information about
the lifetime and/or reaction rates of time-like "particles". The
basis of the partonic phase description is the dynamical
quasiparticle model (DQPM) \cite{Ref26,Ref27} matched to reproduce
lattice QCD results -- including the partonic equation of state --
in thermodynamic equilibrium \cite{Ref24}. In fact, the DQPM allows
a simple and transparent interpretation for thermodynamic quantities
as well as correlators -- measured on the lattice -- by means of
effective strongly interacting partonic quasiparticles with broad
spectral functions. The transition from partonic to hadronic degrees
of freedom is described by covariant transition rates for fusion of
quark-antiquark pairs or three quarks (antiquarks), obeying flavor
current conservation, color neutrality as well as energy-momentum
conservation.

\begin{figure}
\centering
\includegraphics[width=0.5\textwidth]{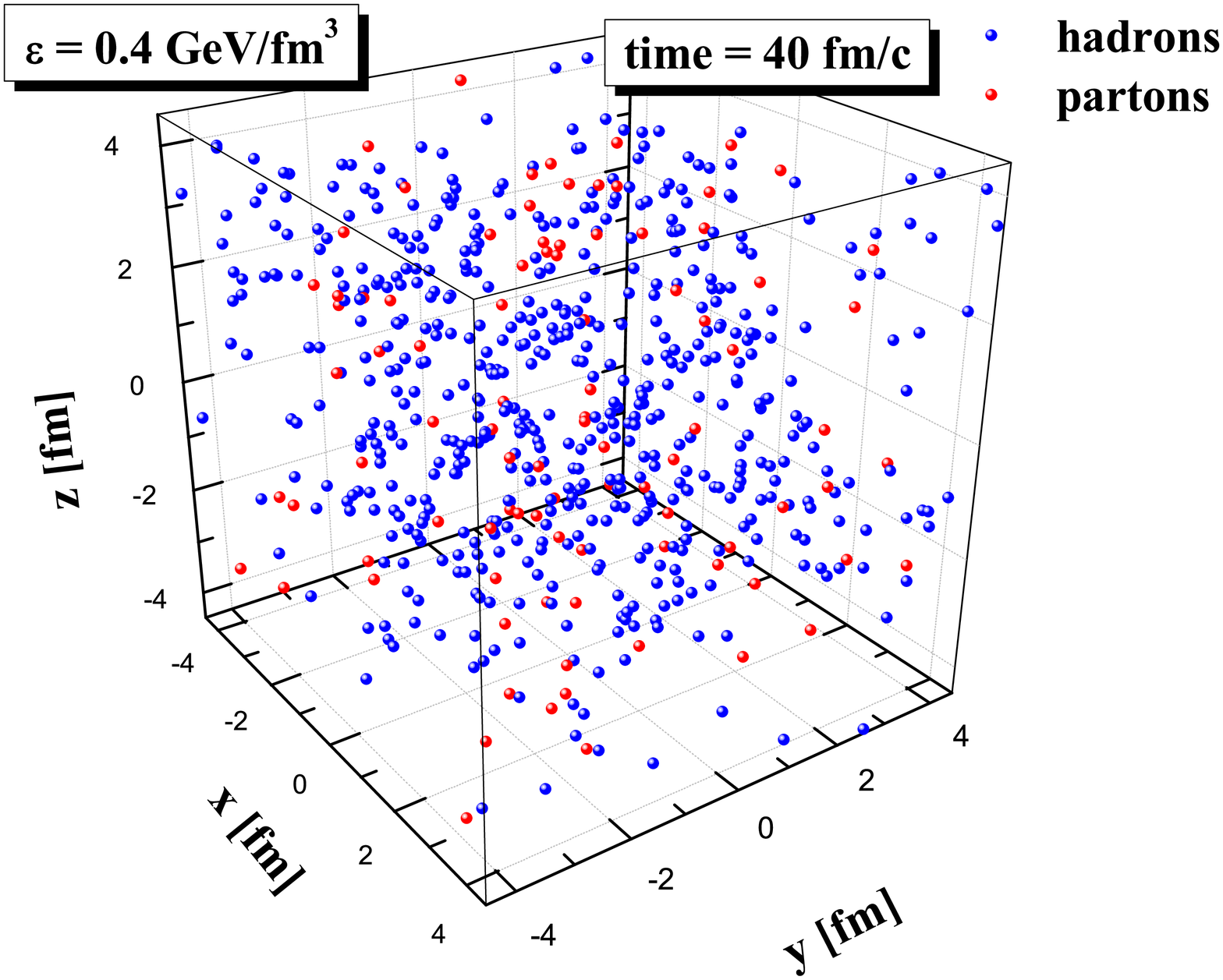}
\caption{Snapshot of the spatial distribution of hadrons (blue) and
partons (red) at an evolution time of 40 fm/c after the systems was
initialized by solely partons at an energy density below critical.}
\label{fig1}
\end{figure}

The `infinite' matter is simulated within a cubic box with periodic
boundary conditions initialized at various values for baryon density
(or chemical potential) and energy density. The size of the box is
fixed to $9^3$ fm$^3$. The initialization is done by populating the
box with light ($u,d,s$) quarks, antiquarks and gluons with random
space positions and the momenta distributed according to the
Fermi-Dirac distribution. The total numbers of the quarks and
antiquarks are chosen so that the system with various desired values
of the energy density $\varepsilon$ (the total energy of the
particles divided by the size of the box) and baryon potential $\mu$
can be studied.

\begin{figure}
\centering
\includegraphics[width=0.5\textwidth]{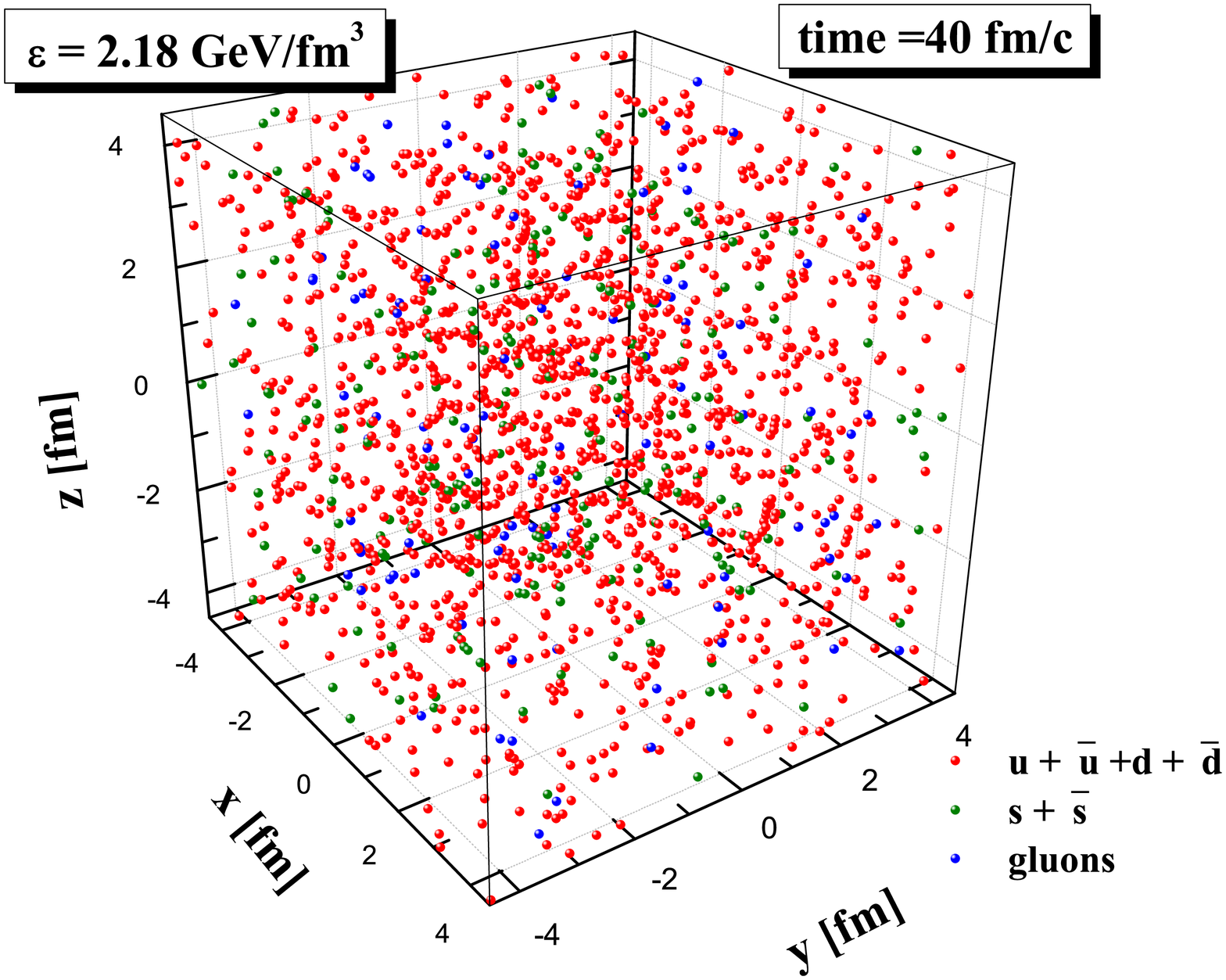}
\caption{Snapshot of the spatial distribution of light quarks and
antiquarks (red), strange quarks and antiquarks (green) and gluons
(blue) at a time of 40 fm/c. } \label{fig2}
\end{figure}

In the course of the subsequent transport evolution of the system by
PHSD, the numbers of gluons, quarks and antiquarks are adjusted
dynamically through the inelastic collisions to equilibrium values,
while the elastic collisions lead to eventual thermalization of all
the particle species (e.g. $u,d,s$ quarks and gluons, if the energy
density in the system is above critical). Please note that if the
energy density in a local cell drops below critical either due to
the local fluctuations or because the system was initialized with a
low enough number of partons, a transition from initial pure
partonic matter to hadronic degrees of freedom occurs dynamically by
interactions.

%------------------------------------------------------------------

\section{Results}
\label{sec:res}

\begin{figure}
\centering
\includegraphics[width=0.5\textwidth]{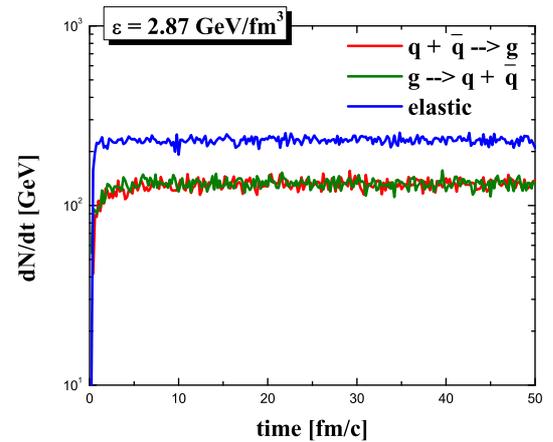}
\caption{The reaction rates for elastic parton scattering (blue),
gluon splitting (green) and flavor neutral $q\bar{q}$ fusion (red)
as a function of time.  } \label{fig3}
\end{figure}

We present in the Fig.~\ref{fig1} a snapshot of the spatial
distribution of hadrons (blue) and partons (red) at an evolution
time of 40 fm/c after the systems was initialized by solely partons
at an energy density of 0.4 GeV/fm$^3$. At this energy density --
slightly below the critical energy density for the deconfinement
phase transition -- most of the partons have already formed hadrons
at this point of the the evolution of the system, but a small
fraction of partons have yet not hadronized. The remaining partons
are approximately in the thermal equilibrium with the hadrons.

\begin{figure}
\centering
\includegraphics[width=0.5\textwidth]{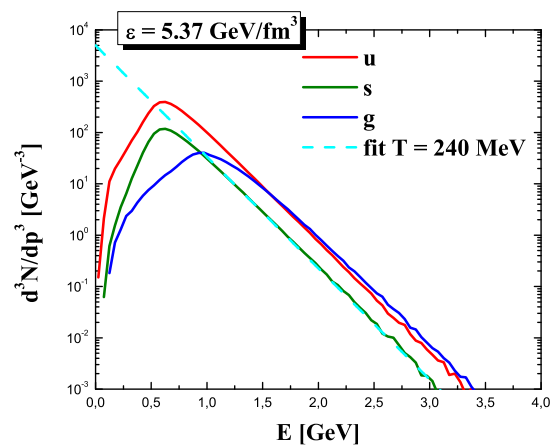}
\caption{The energy spectra for the off-shell u (red) and s quarks
(green) and gluons (blue) in equilibrium for a system initialized at
an energy density of 5.37~GeV/fm$^3$.} \label{fig4}
\end{figure}

In Fig.~\ref{fig2}, we show a snapshot of the systems that has been
initialized at an energy density of 2.18 GeV/fm$^3$, which is
clearly above the critical energy density. One can see in
Fig.~\ref{fig2} the spatial distribution of light quarks and
antiquarks (red), strange quarks and antiquarks (green) and gluons
(blue) at a time of 40 fm/c. At this energy density no hadrons are
seen. The different parton species are in thermal and chemical
equilibrium.

\begin{figure}
\centering
\includegraphics[width=0.5\textwidth]{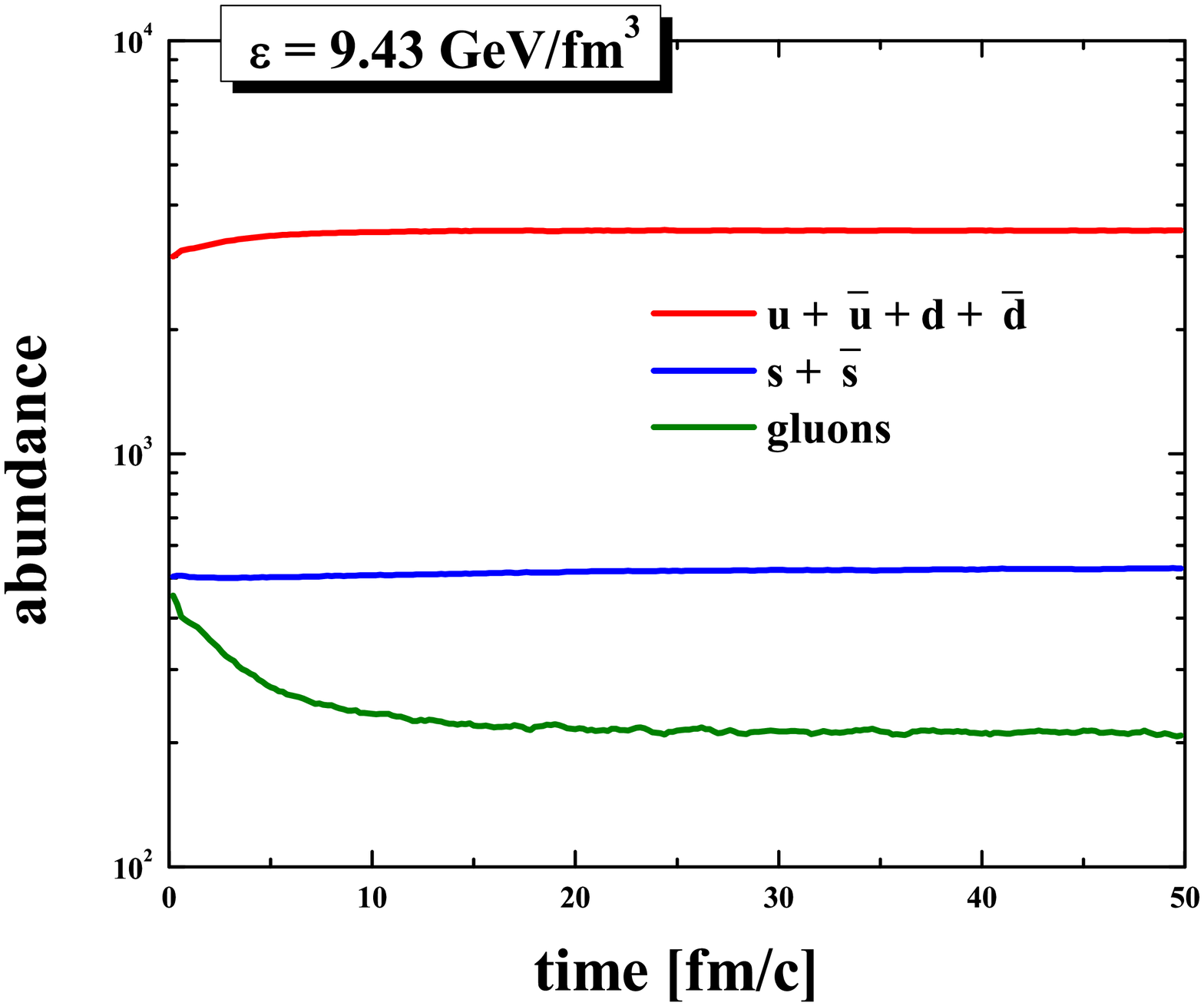}
 \caption{Abundances of the u,d s quarks+antiquarks and
gluons as a function of time for a system initialized at an energy
density of 9.43 GeV/fm$^3$. } \label{fig5}
\end{figure}

After a few fm/c the system -- initialized at an energy density of
2.87 GeV/fm$^3$ -- has achieved chemical and thermal equilibrium,
since the reactions rates are practically constant and obey detailed
balance for gluon splitting and $q\bar{q}$ fusion. This is shown in
Fig.~\ref{fig3}, where the reaction rates for elastic parton
scattering (blue), gluon splitting (green) and flavor neutral
$q\bar{q}$ fusion (red) are presented as a function of time.

\begin{figure}
\centering
\includegraphics[width=0.5\textwidth]{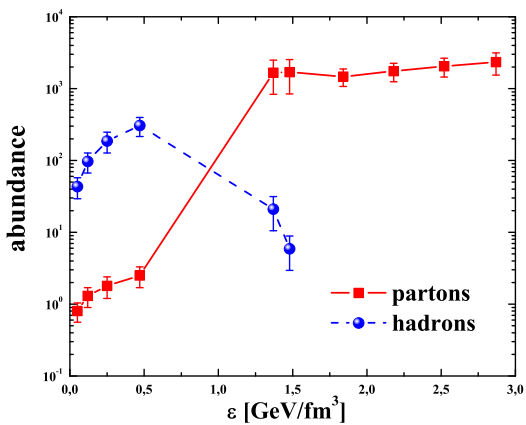}
\caption{The average abundances of hadrons (blue) and partons (red)
in equilibrium as functions of the energy density $\varepsilon$. }
\label{fig6}
\end{figure}

Another indication that the system has achieved the thermal
equilibrium is seen in the distribution of the kinetic energy of the
particles. We show in Fig.~\ref{fig4} the energy spectra for the
off-shell $u$ (red) and $s$ quarks (green) and gluons (blue) in
equilibrium for a system initialized at an energy density of
5.37~GeV/fm$^3$. The spectra may well be described by a Boltzmann
distribution with temperature T=240~MeV in the high energy regime.
The deviations from the Boltzmann distribution at low energy E are
due to the broad spectral functions of the partons.

On the other hand, a sign of the chemical equilibration is the
stabilization of the abundances of the different species. In
Fig.~\ref{fig5}, we show the particle abundances as a function of
time for a system initialized at 9.43~GeV/fm$^3$. One can see that
the chemical equilibration is reached after about 15~fm/c.

It is interesting to observe in Fig.~\ref{fig6} the average
abundances of hadrons (blue) and partons (red) in equilibrium as
functions of the energy density $\varepsilon$. At energy densities
below critical ($\approx$ 0.5~GeV/fm$^3$) the system evolves into a
state, which is dominated by hadrons and has a very small fraction
of partons due to rare fluctuations of local energy density to high
values. At higher energy densities, the system is in a QGP final
state, with a small hadron admixture. At high enough energy (above
approx. 2~GeV/fm$^3$) we find that hadron fraction is negligible. In
the regime of energy densities from 0.48 to 1.3 GeV/fm$^3$ the
calculations have provided so far no stable equilibrium over time
due to large fluctuations between hadronic and partonic
configurations. Further studies are on the way.

%------------------------------------------------------------------
\section{Conclusions}
\label{sec:conclusions}

We have studied the kinetic and chemical equilibration in `infinite'
parton-hadron matter within the Parton-Hadron-String Dynamics
transport approach (PHSD), which is based on a dynamical
quasiparticle model for partons (DQPM) matched to reproduce
lattice-QCD results -- including the partonic equation of state --
in thermodynamic equilibrium.

The `infinite' matter has been simulated within a cubic box with
periodic boundary conditions initialized at different baryon density
(or chemical potential) and energy density.

Depending on the energy density, the system evolved into an ensemble
of either hadrons or partons in chemical and thermal equilibrium.
Abundances of the final particles depend on the energy density. The
temperature of the degrees of freedom in the final state was
measured by fitting the slopes of the Boltzmann-like tails of the
thermal distributions of their kinetic energy.

\section*{Acknowledgements}
V.O. is grateful for the financial support by the Helmholtz ``Quark
Matter" graduate school ``H-QM". O.L., E.B. and M.G. acknowledge the
support by the ``HIC for FAIR" framework of the ``LOEWE" program.

\end{document}